# Electric and Magnetic response in Dielectric Dark states for Low Loss Subwavelength Optical Meta Atoms

*Aditya Jain[1], Parikshit Moitra[2], Thomas Koschny[3], Jason Valentine[4] and Costas M. Soukoulis[3, 5]*

[1] Ames Laboratory—U.S. DOE and Department of Electrical and Computer Engineering, Iowa State University, Ames, IA 50011, USA

[2] Interdisciplinary Materials Science Program, Vanderbilt University, Nashville, Tennessee 37212, USA

[3] Ames Laboratory—U.S. DOE and Department of Physics and Astronomy, Iowa State University, Ames, IA 50011, USA

[4] Department of Mechanical Engineering, Vanderbilt University, Nashville, Tennessee 37212, USA

[5] Institute of Electronic Structure and Lasers (IESL), FORTH, 71110 Heraklion, Crete, Greece

ABSTRACT: Artificially created surfaces or metasurfaces, composed of appropriately shaped subwavelength structures, namely meta-atoms, control light at subwavelength scales. Historically, metasurfaces have used radiating metallic resonators as subwavelength inclusions. However, while resonant optical metasurfaces made from metal have been sufficiently subwavelength in the propagation direction, they are too lossy for many applications. Metasurfaces made out of radiating dielectric resonators have been proposed to solve the loss problem, but are marginally subwavelength at optical frequencies. Here, we design subwavelength resonators made out of non-radiating dielectrics. The resonators are decorated with appropriately placed scatterers, resulting in a meta-atom with an engineered electromagnetic response. As an example, we fabricate and experimentally characterize a metasurface yielding an electric response and theoretically demonstrate a method to obtain a magnetic response at optical frequencies. This design methodology paves the way for metasurfaces that are simultaneously subwavelength and low loss.

Unlike Photonic Crystals, metamaterials derive their properties by modifying the electric and magnetic fields of light, and are free from diffraction[1-8]. The flexibility associated with the geometric control of metamaterials has resulted in fascinating applications like perfect absorbers[9], phase mismatch free non-linear generation[10], magnetic mirrors[11,12], subwavelength cavities[13], zero-index media[14] and slow light devices[15]. To achieve these effects, the essential building blocks of a metamaterial, namely the meta-atoms, must be made sufficiently thin in the direction of propagation of the electromagnetic field. More recently, 2 dimensional versions of metamaterials or metasurfaces[16, 17] have been proposed as an alternative to bulk 3 dimensional metamaterials due to their ease in fabrication, comparatively lower losses and small footprint for on-chip devices. The most popular construction materials for metasurfaces have been radiating metallic antennas[5], although they have scaling issues at optical frequencies[18]. To circumvent this problem, radiating Mie resonances have been proposed in low loss high permittivity particles ($\varepsilon \approx$ 25-1,000) at GHz and lower THz frequencies[19-24]. However, straightforward scaling of this approach to optical frequencies renders isotropic meta-atoms, such as cubes and spheres, marginally subwavelength due to the absence of high index dielectrics[25,26]. Therefore, it is highly desirable to construct low loss meta-atoms made entirely out of dielectrics with modest permittivity ($\varepsilon \approx$ 2-14), whilst still being sufficiently subwavelength in the propagation direction. Recently, researchers have experimented with silicon disks[27, 28], which can be more deeply subwavelength in the direction of propagation at telecommunication frequencies. However, these structures are still not sufficiently thin to compete with their metallic counterparts, especially when utilizing the magnetic response. As a reference, we have simulated the electric and magnetic response in a disk with radiating Mie resonance (see Supporting Figure 1 and Supporting Figure 2). An alternative approach is to access dark modes of the resonators[29], which allows deeper subwavelength thicknesses while still preserving a sharp resonance, an approach that we will address in detail in this paper.

Two main loss channels in metamaterial resonators are present. The first loss channel is the dissipation, which can be reduced by choosing low loss dielectrics. The other loss channel, the radiative loss, can be reduced by suppressing the dipole moment of the resonator. Such non-radiative resonators are more commonly known as 'dark' resonators[30]. This is generally achieved by choosing an appropriate geometry such that the overlap integral of the excited mode and the incident wave is negligible. Another advantage of using dark resonators is that they can be made sufficiently subwavelength in the propagation direction since they don't possess any dipole moment of their own. However, there is a dichotomy in the fact that complete suppression of radiative losses also leads to no metamaterial response. In this Article, we propose a method to solve this problem using non-resonant scatterers. The combination of a non-radiative resonator and an appropriate scatterer results in a hybrid meta-atom. We demonstrate a general recipe to achieve polarization dependent and independent electric response in a single layer metamaterial using such hybrid meta-atoms. Experimental results prove that such resonances can be excited within reasonably subwavelength structures. We then recycle these resonators with another set of scatterers to theoretically demonstrate a meta-atom with a magnetic response as well. All the subsequent discussions concern optical metasurfaces at telecommunication frequencies. We characterize the 2-D array of nanostructures as thin sheets with dimensionless electric or magnetic surface susceptibility ($\chi_{se}^{(d)}$, $\chi_{sm}^{(d)}$) (see supporting information[44, 45]).

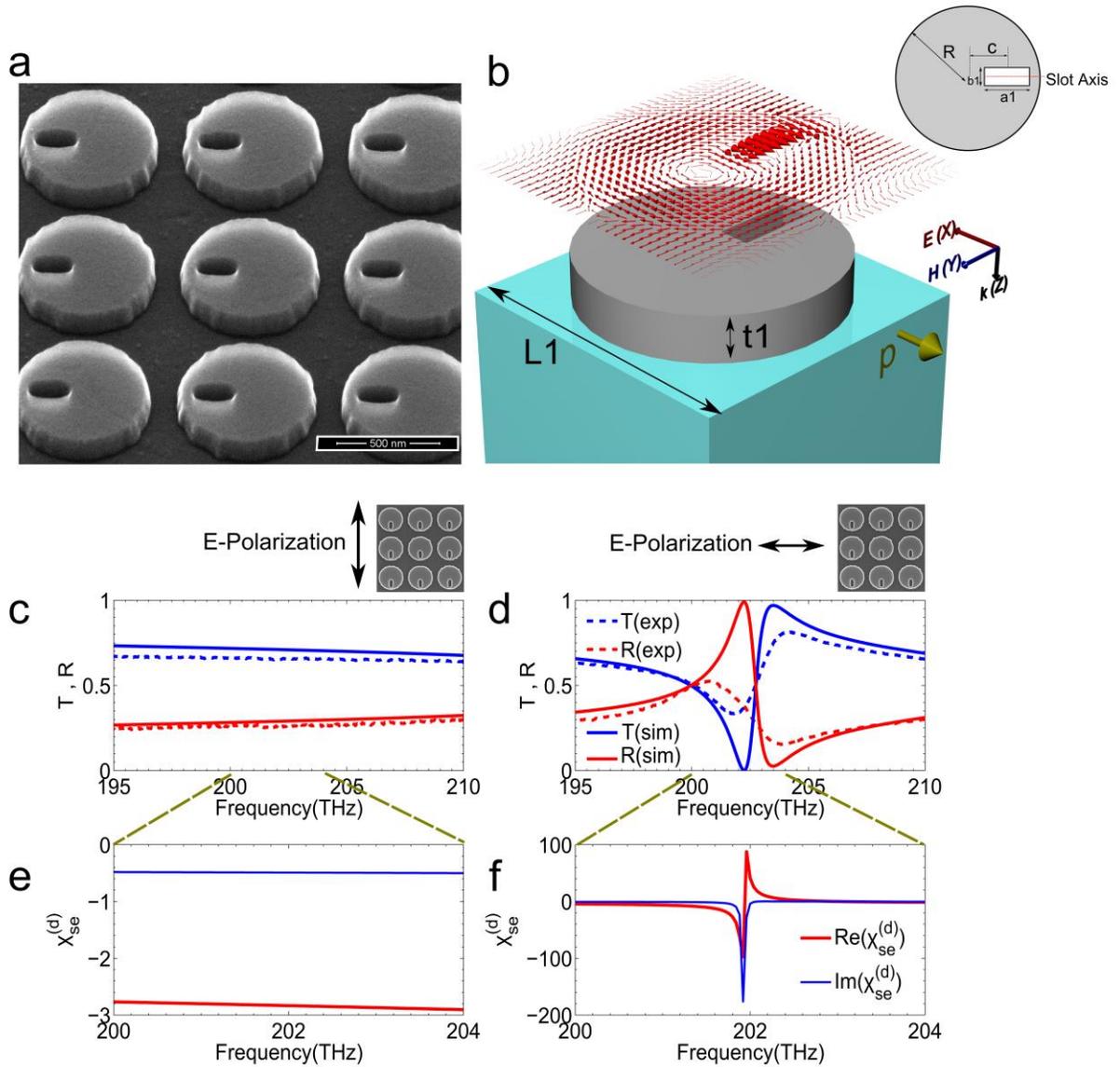

**Figure 1.** Polarization sensitive electric response of a dielectric disk with a rectangular slot. **a**, SEM image of a single layer dielectric disk based metamaterial with an asymmetric through slot. **b**, Simulated unit cell of the proposed design with a dielectric disk made out of silicon (grey) placed on a quartz substrate (blue). E-field at centre XY plane of the disk is projected above the disk (shown in red arrows). Induced dipole moment (shown in green arrow) is along the X direction. Inset: 2-D cross-section of a unit cell with R=315nm, c=165nm, a1=230nm,

b1=90nm t1=115nm, L1=750nm. **c**, Experimental (dashed) and simulated (solid) S-parameter curves for E field polarized along the Y direction. **d**, Experimental (dashed) and Simulated (solid) S-parameter curves for E field polarized along the X direction. **e**, Zoom in of the calculated dimensionless electric surface susceptibility for the E field polarized along the Y direction. No electric response is evoked since the induced electric dipole moment is orthogonal to the incident E-field. **f**, Zoom in of the calculated dimensionless electric surface susceptibility for the E field polarized along the X direction. A strong electric response is evoked since the induced electric dipole is along the incident E-field.

We start our discussion by demonstrating a meta-atom exhibiting an electric response. The first step in the design is to realize a resonator with negligible dipole moment commensurate with the incident wave. The structure is comprised of a silicon disk ($\varepsilon \approx 13.69$, $t_1 =115$ nm; Fig. 1a, b)) with an asymmetrically etched rectangular slot placed on a quartz substrate ($\varepsilon \approx 2.1$). The unit cell is periodically repeated in the X and Y directions to form a metasurface. The incident plane wave has an electric field polarized along the X direction with a propagation vector along the Z direction (Fig. 1b). The dark mode in consideration is the lowest order Mie mode (magnetic dipole mode) in a homogeneous cylindrical disk. The mode frequency is fixed primarily by the radius R of the disk (inset Fig. 1b). The mode has a circulating electric field and doesn't radiate via an electric moment without the slot. The disk however, does radiate via a magnetic moment perpendicular to the plane of the disk (Z direction, Fig. 1b). Nevertheless, the incident H field is along the radial direction of the slab (Y direction, Fig. 1b) and cannot couple to the magnetic moment arising from the Mie mode. Hence, for the purpose of discussion, this mode can be considered dark for the given incident direction. The quality factor (Q-factor) of the dark Mie mode is limited by the radiating magnetic moment and fabrication imperfections (since silicon is lossless around 1.55μm). To generate an electric response function from the homogeneous disk, we place an off-centered slot with its axis along the Y direction (inset Fig. 1b) creating an asymmetry in the structure. The slot serves to scatter light into the dark mode and in turn gets polarized due to the strong fields inside the disk, resulting in an induced electric dipole moment, perpendicular to the axis of the slot (X direction). The projected E-fields (red arrows, Fig. 1b) at the center plane of the disk show high electric field magnitude inside the slot, indicating high residual polarization. For an incident electric field oriented perpendicular to the axis of the slot (X direction, Fig. 1b) coupling is the most efficient and results in a strong electric response at 202 THz (Fig. 1d, 1f). Upon changing the E-field polarization to the Y direction (Fig. 1b), the resonance disappears, resulting in a negligible dipole moment (Fig. 1c, 1e). Experimental results agree well with the simulated results (Fig. 1c, 1d) but the resonances are damped most likely due to fabrication imperfections and surface state absorption in the silicon. The physical thickness of the structure is 115 nm $\approx \lambda_0/13$ ($\lambda_0$ is the free space wavelength at the resonance), which is deep subwavelength for propagation along the Z direction (compare with $\approx \lambda_0/9$ thickness of a radiating electric resonance in a disk, Supporting figure 1). The disks can be made even thinner in the Z direction by stretching them along the lattice direction (X and Y). However, care must be taken to keep the lattice size sufficiently small so as to avoid higher diffraction orders. This scaling is true for all the meta-atoms presented in this work. The methodology presented here is similar to a metallic split ring resonator exhibiting circulating current in a ring with a capacitive gap[1]

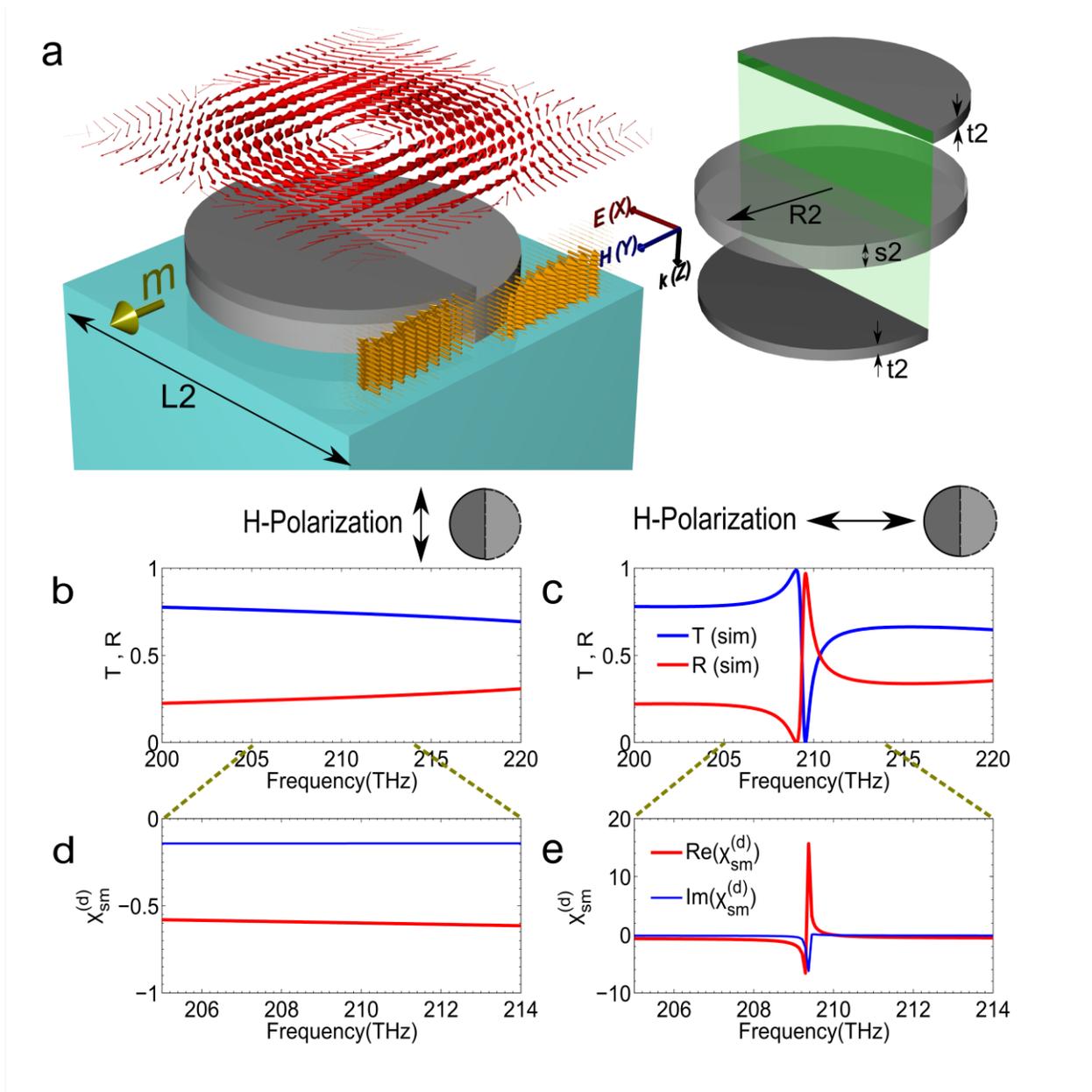

**Figure 1.** Polarization sensitive magnetic response of a dielectric disk with a semi-cylindrical disk scatterer. a, Simulated unit cell of the proposed design with the dielectric disk made out of silicon (grey) placed on quartz substrate (blue). Semi-cylindrical scatterers are placed diagonally across the slab. E-field at the centre XY plane of the disk is projected above the disk (shown in red arrows). Uniform fields indicate zero electric dipole moment. Golden arrows indicate the projected D-field inside the material at the centre YZ plane of the disk. Anti-parallel fields in the scatterers generate a magnetic dipole moment (shown in green arrow). Inset: Exploded view of the magnetic meta-atom, R2=315nm, t2=30nm, s2=55nm, L2=750nm. The semi-transparent green plane indicates the centre plane of the disk b, Simulated S-parameter curves for E field polarized along the Y direction. c, Simulated S-parameter curves for H field polarized along the X direction. d, Zoom in of the calculated effective permeability for the H field polarized along the X direction. No magnetic response is evoked since the induced magnetic dipole moment is orthogonal to the incident H-field. e, Zoom in of the calculated effective permeability and the figure of merit for the H field polarized along the Y direction. A strong magnetic response is evoked since the induced magnetic dipole moment (m) is along the incident H-field.

The second response in consideration is negative permeability. We use the same dark state in the silicon cylindrical disk as presented in the previous section. To generate a magnetic response, we require a loop with circulating current, similar to how magnetic resonances have been implemented in cut-wire pairs and fishnets[7, 8]. To achieve this current loop, we place two scatterers diagonally across the cylindrical disk resulting in the excitation of the dark mode (exploded view inset Fig. 2a). The non-resonant scatterers are shaped as thin semi-cylindrical disks, also made out of silicon. The projected uniform E-field at the center plane of the slab indicates net zero electric dipole moment (red arrows Fig. 2a). This is a direct consequence of the symmetric placement of scatterers about the center axis of the disk (green plane in inset Fig. 2a). In spite of the symmetry, the meta-atom still exhibits a magnetic dipole moment along the Y direction (green arrow Fig. 2a). A scatterer, placed on top of the cylindrical disk (Fig. 2a) gets polarized by the strong near-field of the dark resonator. Similarly, an induced polarization occurs in the scatterer, placed below the cylindrical disk. However, since the scatterers are placed on the opposite half of the dark cylindrical resonator, antiparallel displacement currents arise (see electric displacement field D plotted as golden arrows Fig. 2a). The induced polarization currents in the semi-cylindrical scatterers are in different XY planes, separated by the thickness of dark resonator, thereby generating a circulating current loop with its magnetic moment pointing along the incident H field (Y direction in Fig. 2a). Simulated results indicate magnetic sheet susceptibility at 209 THz accompanied by a narrow line-width resonance, for the incident H-field polarized along the Y direction ($E_x$, $H_y$ in Fig. 2c, 2e). Flipping the incident H-field along the X direction ($E_y$, $H_x$ in Fig. 2b, 2d) results in net-zero response from the meta-atom. The physical thickness of the meta-atom is 115nm $\approx \lambda_0/12.4$, which is again deep subwavelength for wave propagating along Z direction (compare with $\approx \lambda_0/7$ thickness of a radiating magnetic resonance in a disk, supporting figure 2). We therefore have created a low loss dielectric equivalent of a cut-wire pair.

In the previous sections, we have described a general method to excite a purely electric or magnetic response, sensitive only to a single incident polarization. However, to improve the practical applicability of our metamaterial, polarization independent structures are required (invariant response for E field along either X or Y direction). The cylindrical geometry shown in Fig 1 and 2 cannot yield a polarization invariant response without converting a certain fraction of the incident light to cross-polarized transmittance (see Supporting Figure 3 and Supporting Figure 4). This is undesirable for many metamaterial applications. To mitigate this problem, we switch from a cylindrical to a rectangular geometry which allows us to decouple the response in orthogonal directions.

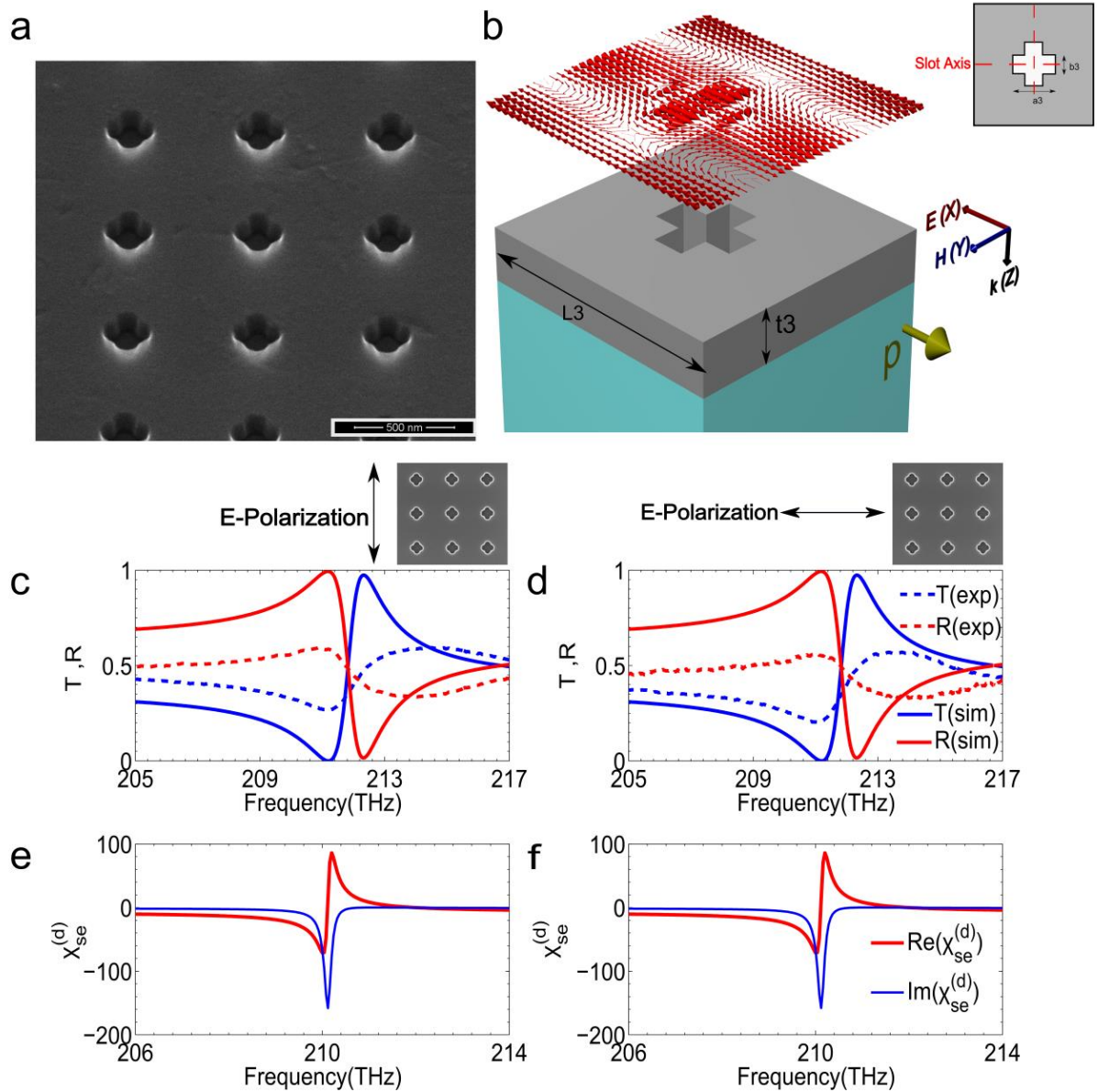

**Figure 2**. Polarization independent electric response of a dielectric slab with a cross scatterer.

**a**, SEM image of a single layer dielectric slab with periodically repeated cross scatterers.

**b**, Simulated unit cell of the proposed design consisting of a dielectric slab made out of silicon (grey) placed on a quartz substrate (blue). The cross slot scatterer is fabricated at the centre of the unit cell by superimposing two orthogonal rectangular slots. Projected E-field at the centre XY plane of the slab (shown in red arrows) showing a strong field inside the cross, inducing an effective electric dipole moment (p shown in green arrow) along the X direction. If the incident E-field is switched to the Y direction (not shown in this figure), similar response is obtained due to the symmetry of the structure. Inset: 2-D cross-section of the unit cell with a3=230nm, b3=90nm t3=115nm, L3=620nm. **c**, Experimental (dashed) and simulated (solid) S-parameter curves for E field polarized along the Y direction. **d**, Experimental (dashed) and Simulated (solid) S-parameter curves for E field polarized along the X direction. **e**, Zoom in of the calculated dimensionless electric surface susceptibility for the E field polarized along the Y direction. A strong electric response is evoked since the induced electric dipole moment is along the incident E-field. **f**, Zoom in of the calculated dimensionless electric surface susceptibility for E field polarized along the X direction. A strong electric response is evoked in the same manner as the response obtained for E-field polarized along the Y direction.

The second dark mode in consideration is the lowest order index guided TE mode in an infinite planar dielectric slab[31]. Guided modes in planar dielectric slab with holes have been studied quite extensively in the past[32]. We revisit such structures in the context of metasurfaces exhibiting electric response.

The fabricated structure consists of a planar silicon thin film ($\varepsilon \approx 13.69$, $t_2 = 115$nm) with a cross slot scatterer, etched through the center of a square lattice (Fig. 3a, inset Fig. 3b). The excited mode profile along the X/Y direction is TE (2, 0) and is fixed by the lattice constant (L3 in Fig. 3b) .This mode cannot be excited from free space without scatterers, due to field symmetry of the mode about XZ plane. Therefore, this mode has no radiation via either electric or magnetic dipole moments and it is perfectly dark (as opposed to the homogeneous disks which possess a radiating magnetic moment perpendicular to the plane of the slab). It is also important to note that the infinite extent of the thin film (along X and Y direction Fig. 3b) allows us to have degenerate versions of the TE (2, 0) mode, whose excitation is dependent on the placement of the scatterers. We appropriately place our scatterers so as to excite only one degenerate mode for a given response. For a given lossless material, the Q-factor of these modes is purely limited by the fabrication imperfections only. To create a polarization independent negative permittivity, we etch a symmetric rectangular slot through the center of the unit cell. To maintain a polarization independent response, we rotate the slot by 90° about the center of the unit cell. Superposition of these two orthogonal slots results in a cross structure (Fig. 3a, inset Fig. 3b). The projected E-field at the center of the slab (red arrows in Fig. 3b) shows the TE (2, 0) mode excited in the resonator. The mode is symmetric ($TE^{symmetric}$ (2, 0)) with respect to the slot axis (inset Fig. 3b). The slot with its axis along the Y direction couples to an incident E-field along the X direction only and vice-versa (see projected E-field Fig. 3b; stronger fields above the slot indicate a residual electric polarization). The simulated and corresponding experimental results for the cross structure clearly show the approximately similar transmittance and reflectance for both X and Y polarized E fields (Fig. 3c, 3d). The calculated dimensionless electric sheet susceptibility from the simulations also remains invariant under the polarization change (Fig. 3e, 3f), thus confirming our approach. The larger damping of the experimentally measured resonances with respect to the simulations can be attributed to increased loss in the silicon arising due to the etch process as well as roughness in the patterned areas. This approach is quite similar to how surface Plasmon polaritons in thin metallic films are coupled to free space via a grating[33].

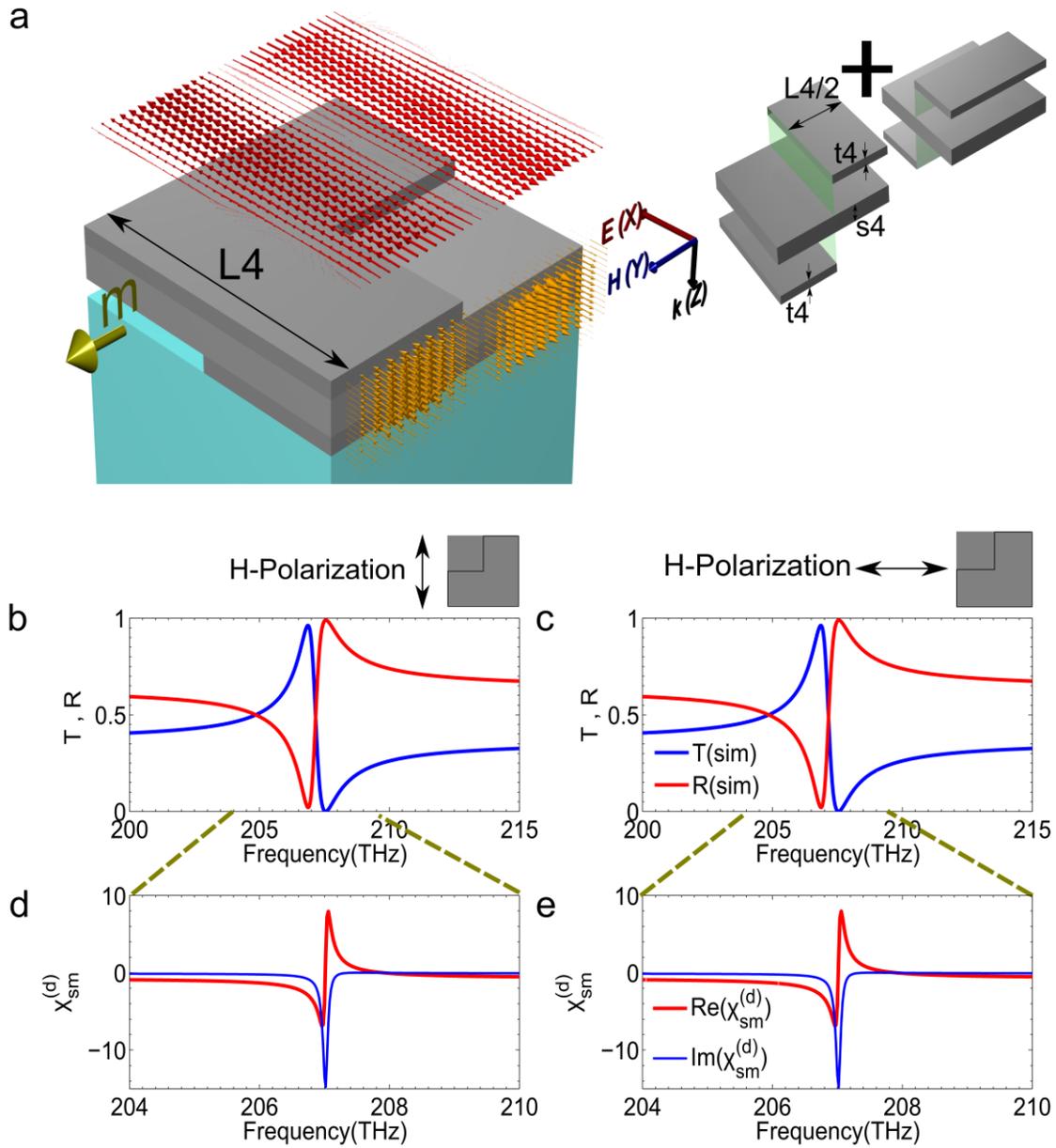

**Figure 3**. Polarization independent magnetic response of a thick dielectric slab with a thin slab scatterer. **a**, Simulated unit cell of the proposed design with a dielectric slab made out of silicon (grey) placed on a quartz substrate (blue). The scatterers are made out of the thin slab, half the width of unit cell, and are placed diagonally across the slab. The E-field at the centre XY plane of the thicker slab is projected above (shown in red arrows). Uniform fields indicate zero electric dipole moment. Golden arrows indicate the projected D-field inside the material at the centre YZ

plane of the disk. Anti-parallel fields in the out of plane scatterers generate a magnetic dipole moment (m shown in green arrow). Inset: Exploded view of the magnetic meta-atom L4=620nm, t4=30nm, s4=55nm. The semi-transparent green plane indicates the centre plane of the thick slab **b**, Simulated S-parameter curves for H field polarized along the Y direction. **c**, Simulated S-parameter curves for H field polarized along the X direction. **d**, Zoom in of the dimensionless magnetic sheet susceptibility for the H field polarized along the Y direction. **e**, Zoom in of the dimensionless magnetic sheet susceptibility for the H field polarized along the X direction.

To excite a polarization independent magnetic response with a planar dielectric slab, we essentially follow the same approach used with a disk resonator. A thin silicon slab, half the width of the unit cell is used as a scatterer. These scatterers are placed diagonally across the thick dielectric slab (inset Fig. 4a). As before, we rotate the scatterers by 90° about the center plane of the unit cell. The original pair of scatterers along with the rotated counterpart is superimposed together to arrive at an L shaped scatterer (Fig. 4a). An important thing to note here is that the excited dark mode differs from the polarization independent electric response case (compare red E-field arrows in Fig. 3b with red E-field arrows in Fig. 4a). The mode profile in the Y direction is anti-symmetric (TE $^{\text{anti-symmetric}}$ (2, 0)) with respect to the slot axis (semi-transparent green plane in inset Fig. 4a). Only the TE $^{\text{anti-symmetric}}$ (2, 0) mode contributes to the response for the current arrangement of scatterers. Simulated D-fields (golden arrows in Fig. 4a) indicate that the current flow in the top and bottom scatterer is antiparallel. As these currents lie along different XY planes, a magnetic moment $m_y$ (green Arrow, Fig. 4a) is generated for an incident $E_x$ field and vice-versa. The simulation results and the retrieved dimensionless magnetic sheet susceptibility (Fig. 4b, 4d and Fig. 4c, 4e) clearly indicate an equivalent magnetic response for E field polarized either along the X or Y direction.

We have theoretically and experimentally demonstrated a method to design subwavelength dielectric metamaterials by splitting the response into two components. The first component consists of a non-radiative or dark resonator, which stores the major fraction of the electromagnetic energy. The second component is the non-resonant scatterer, which imparts this dark-resonator its desired response. The response can be easily altered by changing the geometry of the scatterers. Thus, other responses like chirality or non-linearity can be obtained by a judicious choice of the scatterers[34, 35]. This approach is significantly different from regular metamaterial structures where a single resonator stores and dissipates energy due to finite polarizability of the structure. In this work, we decouple the response of the resonator from its geometry, which imparts greater versatility to the design process. The magnitude of the response can be tuned by simply changing the coupling between the dark mode resonator and the scatterer (changing thickness or radius/width of scatterer). The principles presented in this article can be extended to any structure with negligible dipole moment. Approaches like conformal mapping might enable more complex geometries with deeper subwavelength meta-atoms[36, 37]. The electromagnetic response of existing dark metallic meta-atoms[38, 39] can also be altered using this approach, provided loss can be compensated by gain[39-42]. This method is not limited to classical resonators and atomic transitions can also be used[43], if an appropriate scatterer is available. The reduction of dissipation and compact dimensions in dielectric metasurfaces is very desirable for applications involving a normally incident beam, like slow light devices, modulators, non-linear frequency conversion, polarization convertors or optical isolation. Thus, our work offers a method to enable these applications at much higher frequencies and with much more compact geometries.

ASSOCIATED CONTENT

**Supporting Information**. Fabrication of Disk and Slab , Optical Characterization, susceptibility calculations ,Electric and Magnetic response of a homogeneous disk with radiating mode , cross polarization properties of the structure for rotated slot axis, cross polarization properties of the structure with two orthogonal slots .This material is available free of charge via the Internet at http://pubs.acs.org.

AUTHOR INFORMATION

**Corresponding Author**

*Name: Aditya Jain *Email: ajain17@iastate.edu.

**Author Contributions**

A.J., T.K. and C.M.S conceived the idea and A.J. conducted the numerical simulations and calculations. P.M. and J.V. fabricated and experimentally characterized the metamaterials. A.J wrote the manuscript with contributions from all authors. All authors have given approval to the final version of the manuscript.

**Notes**

The authors declare no competing financial interests.

ACKNOWLEDGMENT

Work at Ames Laboratory was partially supported by the U.S. Department of Energy, Office of Basic Energy Science, Division of Materials Sciences and Engineering, Contract No. DE-DE-AC02-07CH11358 (theory), and by the US office of Naval Research, Award No. N00014-14-1-0474 (simulation) and Award No. N00014-14-1-0475 (experiments).

Table of Contents Graphic

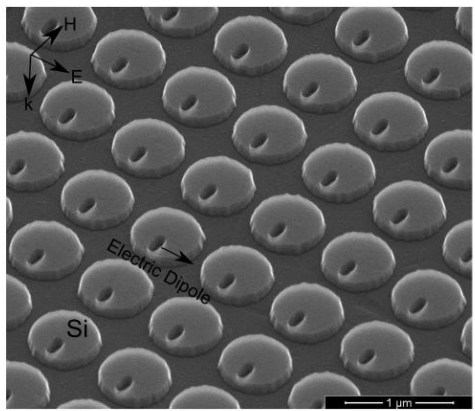
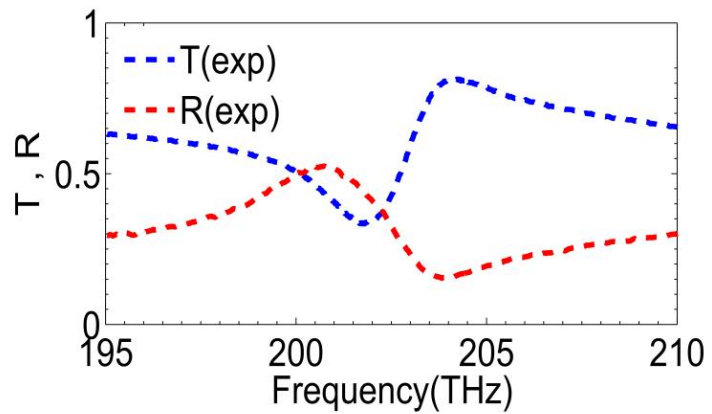